# EVOLUTION OF THE SPS POWER CONVERTER CONTROLS TOWARDS THE LHC ERA

J.C.L. Brazier, Brazier Systems & Consultants Ltd, UK
A. Dinius, CERN, Geneva, Switzerland, P. Semanaz, B & S International, France

Abstract

By the end of the nineties, the power converter control system (Mugef) of the CERN proton accelerator (SPS) had undergone a complete modernization. This resulted in newly developed hardware for function generation, measurement and I/O in a VME environment, under the LynxOS real-time operating system. This has provided a platform on which extensions can be developed for future operation in the Large Hadron Collider (LHC) era. This paper describes some of these extensions, in particular a fast Surveillance and Interlock system for monitoring the power converter output currents. This will be mandatory for the safe operation of the SPS transfer lines TI2 & TI8 to LHC and for similar applications in the future. The strategies employed to cope with various failure modes of the power converters and the timely activation of the interlock are outlined. The new SPS controls infrastructure now under development, will give rise to new modes of operation for the Mugef systems. Integration with the proposed middleware must be undertaken in a structured evolution, while retaining compatibility with the current usage.

## 1 BACKGROUND

Magnet circuits in CERN's SPS are driven by power converters. These power converters are controlled by equipment known as a Mugef [1] that:

- Turns the power converter on and off and reads back its state.
- Using an analogue to digital converter (ADC), digitises the analogue voltage (±10V) from the output current transducer (DCCT) every millisecond.
- Produces a (±10V) reference voltage for the power converter to generate the magnet current, using a digital to analogue converter (DAC). The output of the DAC follows a pre-loaded function.

One Mugef crate can control up to 64 power converters and connects to the control system via a proprietary command/response middleware (SL-Equip).

### 1.1 Mugef Description

The present Mugef is described in [1]. It consists of a VME chassis with a PowerPC-603e. The controlling software runs under the LynxOS real time operating system. Special purpose cards perform the reference voltage function generation and the analogue acquisition of the reference output and DCCT output. A single card provides on/off/reset and status read-back for all the power converters. A CERN standard timing card provides synchronisation and SPS machine timing events. A System Arbiter Controller (SAC) card completes the hardware.

The controlling software consists of a set of processes connected with message queues and shared memory segments. There is a process controlling each of the hardware elements (function handling, acquisition, status control, etc.) which includes the hardware driving, as all the cards are directly mapped into the process' memory. A timing process translates machine timing events into internal action triggers. These are interpreted by an action sequencer that maintains a list of actions (loading functions, doing measurements etc.) Finally, there is a process that accepts commands from the rest of the control system and initiates immediate actions ('*Now*'). This software is general purpose and has been ported to other applications within the SPS control system [2].

### 1.2 Operational Experience

Some 24 Mugef systems were brought into operation in April 1998 and have worked well since. Reliability has been good, with several months between reboots.

One of the initial design criteria was that the system should be amenable to extensions and modifications. Several of these have been absorbed successfully, most notably, the ability to accept dynamic corrections to reference output function data. This was included for experiments in closed-loop control of the beam tune [3]. This entailed receiving a stream of correction data via UDP over a private network from beam position equipment around the ring. These presaged the closed-loop control needed for LHC.

The following chapters will detail two crucial extensions, namely a dynamic surveillance system

(known as Channel-64) and the modification of the controlling software to mate with the proposed SPS control system middleware.

## 2 CHANNEL-64

Surveillance of the power converters in operation has been, in the most part, the province of the control system application programs. These perform both occasional analogue measurements of the output current and repeated requests for power converter status. Whilst this is adequate for many cases, particular circumstances such as beam ejection require rigorous testing of the magnet current at the time of ejection, on a cycle-by-cycle basis. Here it is necessary to ensure that all the critical power converters in the transfer line are working *within their expected tolerance* just before the beam is extracted. Implicitly, this also means that the power converter is not in fault and therefore a status check is not needed. If any of the critical tolerances are exceeded, then extraction must be prevented, otherwise material damage to the vacuum chamber and magnets could occur.

### 2.1 Channel-6 Heritage

The Channel-6 system, (named after the $6^{th}$ interlock chain in the SPS control room), was designed in the mid 80's, using discrete analogue electronic components to provide the above functionality. A total of 18 power converters had to be monitored in two buildings. Remote control of the Channel-6 system was done via standard SPS multiplex and associated software.

The system verified that the monitored power converters had reached their nominal current 20 ms before the extraction should take place. On a per power converter basis, two comparisons using analogue comparators were made, just prior to extraction. Using the timing system to trigger the process, the DCCT output signal was compared with a reference voltage supplied by the Mugef system. Any difference exceeding the set tolerance then opens an interlock chain, thus disabling the extraction kicker magnet power converter. This prevents extraction of the beam.

Since this hardware was reaching its end of life and the same functionality is needed for the LHC transfer lines, the opportunity existed to provide the same functionality, using an extension of the Mugef system. Specifically, advantage could be taken of the fact that:

- The Mugef could digitally compare the actual ADC output values with the theoretical function values, rather than using a hardware-synthesized reference and analogue comparison.
- The analogue acquisition memory already contains data, at the milli-second rate, prior to the comparison.
- A choice of which combination of power converters to include is fully variable up to all 64 available.
- Different comparison algorithms can be used in different circumstances.
- The precision of the comparison can be increased from 2% to 0.1%

These advantages lead to a very cost effective solution, both for upgrading the original Channel-6 equipment and also for the new transfer lines to LHC.

With any safety/interlock system, great care needs to be taken to ensure the integrity of the system. Permissioning must be positive, ie. the system has to enable the output only after each good comparison. The design must ensure that a previous GO situation is not erroneously replicated to succeeding cycles. This is taken care of by both software and hardware clearing the permission at the start of each machine cycle.

### 2.3 Hardware Needed.

No additional hardware is needed if the power converters monitored by the Channel-64 system are only sending a message to the Central Alarm Server (CAS) when the set tolerances are not met. However for beam extraction, some additional hardware is needed. However, instead of adding an I/O board, the existing SAC board is used for performing the necessary I/O tasks for the Channel-64 hardware. Since the internal memory on the SAC board is not used, a small daughter board has been developed to replace the memory chips. Outputs are connected via a flat cable to the Channel-64 display and interlock chain chassis. A maximum of sixteen individual interlock chain chassis can be driven from the SAC board.

### 2.4 Software Additions

The surveillance is triggered by a machine event occurring >= 10 mS prior to the required output action. Typically, some 3-10 mS of earlier data has been used. Using a tolerance table, pre-loaded with allowable values for each selected channel, a comparison is made with the measured data. Timing test have revealed that the entire process takes approximately 4 mS for all 64 channels.

A result bit-mask is returned of all the channels out of tolerance. This is sent to the hardware in the form of go/no-go and the appropriate output is/is-not enabled.

### 2.4.1 Data Filtering Strategy Routines

The original Channel-6 system simply averaged over time and compared the result with the tolerance using analog comparators. Whilst this is acceptable with stable ADC values, it will give erroneous results on a ramp. In this case, point by point comparison with the reference may be better.

In certain cases, power converter ripple and noise may be a problem. Simple averaging will mask the noise, whereas point comparisons will not. A variation to the point comparison strategy is not to allow single spikes to affect the result. This is accomplished using a median filter. Ripple outside a given tolerance could also be detected by yet another data treatment.

Because different strategies are valid in different circumstances, a mechanism has been incorporated to allow different *data treatment subroutines* to be incorporated with the minimum of effort.

Preliminary tests have proven the validity of the system using the strategies outlined above. Detailed results can be found in [4].

### 2.5 Software Only Solutions

Since the software is available on all the Mugef systems, regardless of whether they have the attached hardware, this becomes a useful tool for monitoring power converters in operation. To this end, several new commands have been added to allow the surveillance results to be logged or read directly.

### 2.5.1 Alternative To on/off Status Checking

In general, status checking only confirms that the required power converters are switched on. It does not confirm that their output current is within tolerance. By comparison the new surveillance provided by Channel-64 implicitly provides *all* this information. Therefore, many application programs could be simplified/speeded up by augmenting traditional status checking with this new method.

## 3 NEW SPS CONTROL SYSTEM

The current power converter application programs interact with Mugef via an infrastructure based on the SL-Equip call, a simple, single-user, synchronous, command response protocol. This may be superseded in the near future by a new middleware-based control software, with which the Mugef must necessarily integrate.

This middleware, is part of a project to upgrade of the SPS software[5] to turn the SPS into a multi-cycling injector for LHC. This works by defining management *contracts* for cycle, settings and status control. A *'Device Server'* in the equipment subscribes to as many contracts as it can implement. Callback routines in the Device Server are activated as part of the contract, for example on the addition/removal of new machine cycles.

The new control system and device servers are being developed now but, will have to be implemented on a phased basis. During this transition, both the new and old models of control will have to co-exist. To enable this, a 'Bridge Server' is being constructed which will respond to the new middleware calls and generate the appropriate SL-Equip calls to the Mugef. Initially, this will implement the '*state contract*', which allows a power converter to be turned on and off. Other contracts, like 'settings' and 'cycle', will be added in due course.

When all the legacy applications have disappeared, the bridge can be disposed of and its functionality absorbed into the basic structure. Clearly, this will be a major software change (approx. 40%) entailing a complete rewrite of the Mugef front end. However, the original design of the Mugef software allows such changes to be made with relative ease, as previous examples have shown.

## 4 CONCLUSION

The hardware and software employed in the Mugef was designed from the start to be flexible enough to incorporate changes, as requirements for power converter control evolved. During its lifetime, the whole SPS control system model will also change. This paper shows the system fulfilling such requirements and rising to the new challenges of the LHC era.